\documentclass[aps,prb,twocolumn]{revtex4-2}

\pdfoutput=1

\usepackage{amsfonts,amsmath,amssymb}

\usepackage{graphicx}
\graphicspath{{graphics/}} 

\usepackage{xcolor}
\usepackage[colorlinks,citecolor=blue,linkcolor=blue,urlcolor=blue]{hyperref}

\usepackage{lmodern} 
\usepackage[T1]{fontenc}
\usepackage[utf8]{inputenc} 
\usepackage{nicefrac}
\usepackage{textcomp}
\usepackage{braket}
\usepackage{slashed}
\usepackage{tensor}
\usepackage[subnum]{cases}
\usepackage{mathtools}
\usepackage{bm}

\newcommand{\h}{\hbar}

\begin{document}

\title{Strong quantum interaction between excitons bound by cavity photon exchange} 

\author{ Miguel S. Oliveira}
\author{Cristiano Ciuti}
\affiliation{Université Paris Cité, CNRS, Matériaux et Phénomènes Quantiques, 75013 Paris, France}

\begin{abstract}
We theoretically predict the interaction between polaritonic excitations arising from the coupling of a cavity photon mode with bound to continuum intersubband transitions in a doped quantum well. The resulting exciton bound by photon exchange, recently demonstrated experimentally, exhibits a binding energy that can be continuously tuned by varying the cavity frequency. We show that polariton–polariton interactions, originating from both Coulomb interactions and Pauli blocking, can be dramatically enhanced by reducing the exciton binding energy, thereby increasing the effective Bohr radius along the growth (vertical) direction. This regime is reminiscent of Rydberg atoms, where weak binding leads to strong quantum interactions. Our predictions indicate that this physics can give rise to giant quantum optical nonlinearities in the mid and far infrared, a spectral region that remains largely unexplored in quantum optics and offers exciting opportunities for both fundamental studies and applications.
\end{abstract}

\maketitle

\section{Introduction}
Polaritons are hybrid quasiparticles arising from the strong coupling between light and matter. These quasiparticles can be realized using various types of electronic excitations, including excitons (Coulomb-bound electron-hole pairs) in semiconductors and 2D van der Waals materials, as well as Rydberg atoms. In the case of exciton-polaritons, polariton-polariton interactions have enabled remarkable phenomena such as Bose-Einstein condensation, superfluidity, and nonlinear optical effects~\cite{RMP2013}, with growing interest now focused on the emergence of strongly correlated regimes~\cite{Bloch2022strongly}. A Rydberg atom is, by definition, an atom excited to a bound state with energy close to the ionization continuum, characterized by a Bohr radius much larger than that of the ground state. This feature enables giant quantum optical nonlinearities and the formation of strongly correlated quantum states~\cite{SaffmanRMP2010}. In atomic systems, Rydberg polaritons exhibit interactions so strong that they can induce quantum blockade, whereby the resonant injection of a single polariton prevents the absorption of a second one~\cite{Gorshkov2011,Jia2018}.
Recently, there has been growing interest in extending the Rydberg concept to excitons, with the goal of harnessing enhanced quantum optical nonlinearities~\cite{Gu2021,Coriolano2022}. In this context, one considers highly excited states of Coulomb-bound electron-hole pairs with energies approaching the ionization continuum.

Intersubband polaritons~\cite{Dini2003,Ciuti2005,Todorov2010,Colombelli2015}, based on doped semiconductor quantum wells, offer a fundamentally distinct platform operating in the infrared and THz spectral ranges, as opposed to the visible regime of exciton-polaritons. These polaritons result from the strong coupling between a cavity mode and the collective intersubband excitation of electrons in a quantum well (intersubband plasmons), where energy conduction subbands are quantized due to confinement. A recently discovered class of intersubband polaritons arises from the strong coupling between a cavity photon and bound-to-continuum intersubband transitions in a doped quantum well~\cite{Cortese2019,Cortese2021}. This coupling gives rise to an exciton-like state bound via photon exchange. While polariton-polariton interactions have been theoretically investigated for conventional polaritons based on bound-to-bound intersubband transitions~\cite{Nguyen-The2013}, and their nonlinear effects have been recently demonstrated experimentally~\cite{Knorr2022}, the interactions involving these unconventional polaritons originating from bound-to-continuum intersubband transitions remain unexplored.

In this article, we theoretically demonstrate that the interactions between polaritons arising from the strong coupling of a cavity photon mode to bound-to-continuum intersubband transitions in a doped quantum well can be dramatically enhanced by controllably reducing their binding energy through the cavity-mode frequency detuning. The paper is organized as follows. In Sec. \ref{sec:Framework}, we present the theoretical framework based on the quantum light–matter interaction and Coulomb Hamiltonian for a doped semiconductor quantum well coupled to a cavity photon mode. In Sec. \ref{sec:bound}, we develop the theoretical treatment of these non-conventional polaritons and evaluate their interactions, showing that their quantum nonlinearities are significantly enhanced compared to standard intersubband polaritons originating from bound-to-bound intersubband transitions. Finally, conclusions and perspectives are provided in Sec. \ref{sec:conclusions}. Technical details concerning the continuum states and the photonic spectral function are given in Appendix \ref{App:continuum} and Appendix \ref{App:photonic}, respectively. 

\section{Theoretical framework}
\label{sec:Framework}
In the following, we will consider the quantum Hamiltonian of a two-dimensional electron gas in a quantum well, including Coulomb interaction and the coupling to a cavity quantum electromagnetic mode, namely 
\begin{equation}
\hat{H} =\sum_{\lambda,\textbf{k}}\epsilon_{\lambda\textbf{k}}\hat{c}^\dagger_{\lambda\textbf{k}} \hat{c}_{\lambda\textbf{k}}, + \hat{H}_{\text{Coul}} + \hbar  \omega_c \hat{a}^\dagger \hat{a} + \hat{H}_{\text{el-cav}}\,  ,
    \label{H_cav}
\end{equation}
where the operator $\hat{c}^\dagger_{\lambda\textbf{k}}$($\hat{c}_{\lambda,\textbf{k}}$) is the fermionic operator creating (annihilating) an electron in the single particle eigenstate belonging to the conduction quantum well subband $\lambda$ with energy $\epsilon_{\lambda\textbf{k}}$ and  in-plane wavevector $\textbf{k}$.  Note that for simplicity we will omit the spin index, because the light-matter coupling conserves the spin and the system is spin degenerate.  The corresponding orbital wavefunctions are $\langle\textbf{r}|\lambda,\textbf{k}\rangle=\varphi_\lambda(z)e^{i\textbf{k}\cdot \textbf{r}_\parallel}/\sqrt{S}$, where $S$ is the transverse area. The second term $\hat{H}_{\text{Coul}}$ describes the Coulomb electron-electron repulsion:
\begin{equation}
    \hat{H}_{\text{Coul}}=\frac{1}{2}\sum_{\substack{{\bf k},{\bf k'},{\bf q}\\
\mu,\mu',\nu,\nu'
}
}V_{q}^{\mu\nu\nu'\mu'}\hat{c}_{\mu,{\bf k+q}}^{\dagger} \hat{c}_{\nu,{\bf k'-q}}^{\dagger} \hat{c}_{\nu',{\bf k'}} \hat{c}_{\mu'{\bf k}},\label{Eq:Coulomb}
\end{equation}
where the Coulomb matrix elements depend on integrals involving the envelope wave functions $\varphi_\lambda(z)$, namely
\begin{align}
V_{q}^{\mu\nu\nu'\mu'} & =\frac{e^{2}}{2S\epsilon_{0}\epsilon_{r}q}I_{q}^{\mu\nu\nu'\mu'},\label{VCoulomb}\\
I_{q}^{\mu\nu\nu'\mu'} & =\int dzdz'\varphi_{\mu}(z)\varphi_{\nu}(z')\varphi_{\nu'}(z')\varphi_{\mu'}(z)e^{-q\left|z'-z\right|}\label{Coherence-func}.
\end{align}
 
The bosonic operator $\hat{a}^\dagger (\hat{a}) $ creates (destroys) a photon in the considered cavity electromagnetic mode with frequency  $\omega_c $. Finally, $\hat{H}_{\text{el-cav}}$ describes the cavity coupling to the electron gas.  

\begin{figure*}[t!]
\begin{centering}
\includegraphics[scale=0.3]{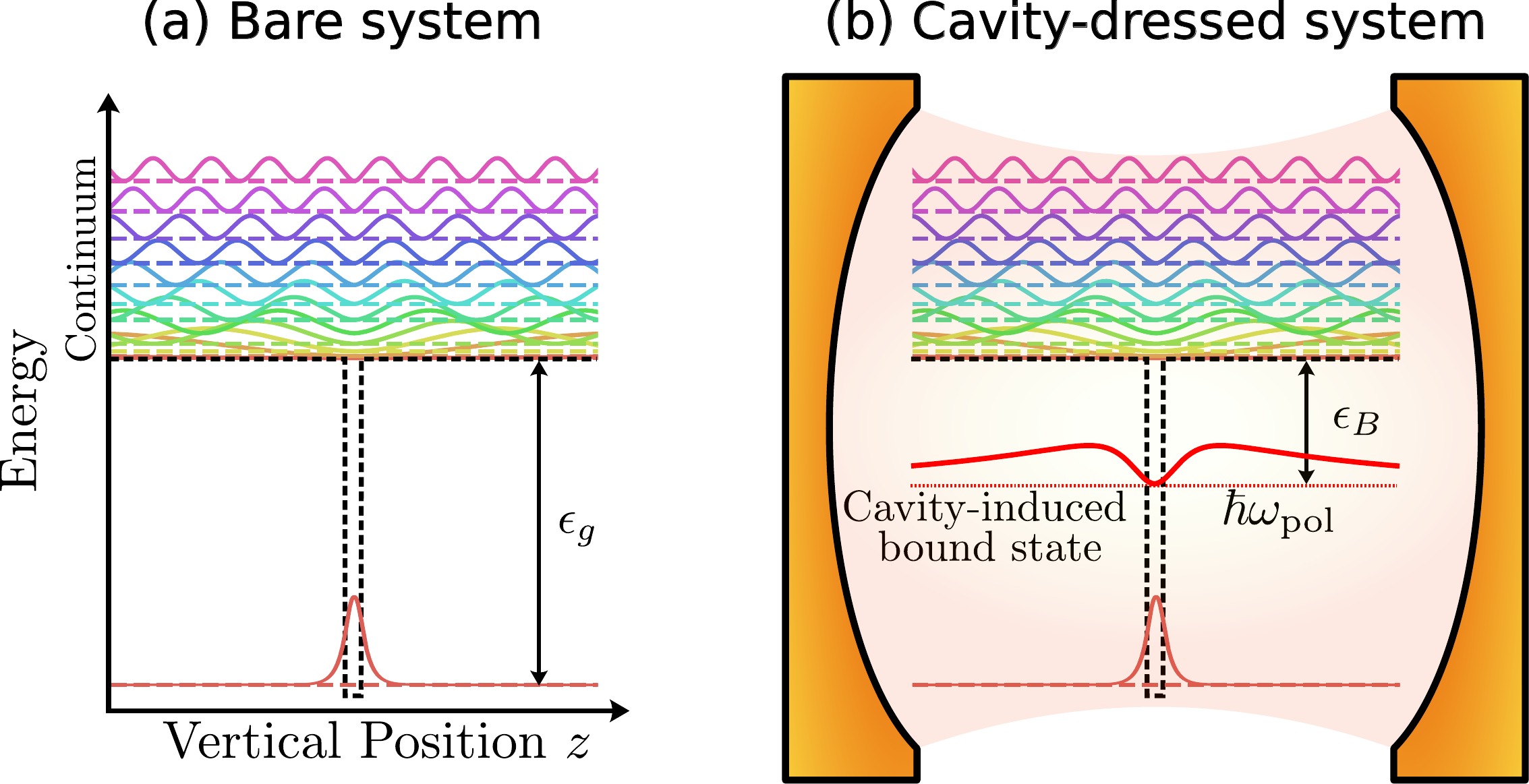}
\par\end{centering}
\caption{Schematic representation of the single-particle electronic spectrum of a quantum well with one quantum-confined conduction subband separated from the continuum by an ionization energy gap $\epsilon_g$.
(a) Square modulus of the envelope functions of the quantum-well electron eigenstates along the $z$ direction, in the absence of a cavity mode. The quantum-well potential and energy levels are indicated by the black dashed line.
(b) Cavity-dressed system. The coupling between the bound-to-continuum transitions and the cavity mode gives rise to a polariton bound state, whose electron-component probability density is shown as a red solid line. The degree of localization of this electron component depends sensitively on the polariton binding energy, defined as the difference between the polariton excitation energy $\hbar\omega_\text{pol}$ and the gap energy $\epsilon_g$. This binding energy can be tuned by adjusting the cavity frequency, as discussed in the main text.\label{fig:qw-states}}
\end{figure*}

As depicted in the left panel of Fig. \ref{fig:qw-states}, in the following we will consider the situation where the quantum well potential is shallow enough to host only one quantum confined subband (corresponding to $\lambda=1$). The other subbands are not quantum confined and form the continuum of the quantum well single-particle energy spectrum. 
For the light-matter interaction, we will consider the quantum paramagnetic interaction in the rotating-wave approximation, namely:
\begin{equation}
    \hat{H}_{\text{el-cav}}=\sum_{\mu,\bf{k}}\hbar \chi_{\mu} ( \hat{c}_{ 1\textbf{k}}^\dagger \hat{c}_{\mu \textbf{k}}\hat{a}^\dagger+ \hat{c}_{\mu\textbf{k}}^\dagger \hat{c}_{ 1\textbf{k}} \hat{a}) \,.
    \label{H_el-cav-fermi}
\end{equation}
Here, $\mu$ labels the continuum subband states. Note that in this article we focus on the strong light–matter coupling regime, but not on the ultrastrong coupling regime, where non-rotating-wave contributions to the light–matter interaction and diamagnetic corrections become relevant~\cite{Ciuti2005}. Neglecting such effects, the ground state of the system is a Fermi sea of electrons occupying the first conduction subband, multiplied by the photon vacuum $\vert 0 \rangle$. The Fermi sea state is given by $|F\rangle \equiv \prod_{k<k_F} \hat{c}_{1\textbf{k}}^\dagger |0\rangle_{el}$, where $|0\rangle_{el}$ denotes the electron vacuum state of the conduction band.

The light-matter coupling can be recast as
\begin{equation}
    \hat{H}_{\text{el-cav}}=\sum_{\mu}\hbar \Omega_{\mu} (\hat{a}^\dagger \hat{b}_{\mu}+\hat{b}_{\mu}^\dagger \hat{a}),
    \label{H_el-cav}
\end{equation}
where we introduced the collective bound-to-continuum bright excitation operators,
\begin{equation}
    \hat{b}^\dagger_{\mu}=\frac{1}{\sqrt{N}}\sum_{|\textbf{k}|<k_F}\hat{c}^\dagger_{\mu\textbf{k}} \hat{c}_{1\textbf{k}},
\end{equation}
where $N$ is the total number of electrons occupying the lowest energy conduction subband. The coupling between the cavity photon mode and such excitations is given by the collective vacuum Rabi frequency $\Omega_{\mu}$
\begin{equation}
    \hbar\Omega_{\mu}=\sqrt{\frac{\hbar\omega_c n_e}{2\epsilon L_{\text{cav}}}}ez_{1\mu},
\label{RabiFreq}
\end{equation}
where $n_e$, $\epsilon$ and $z_{1\mu}$ are, respectively, the electronic density per unit area, the semiconductor's dielectric constant and the electron dipole matrix element corresponding to the transition from the quantum confined subband ($\lambda = 1$) state to a continuum state labeled by $\mu$.  

These bright intersubband excitations obey bosonic commutation relations only approximately~\cite{DeLiberato2009,Nguyen-The2013}; however, the deviation from perfect bosonicity is an operator whose matrix elements are of order $N^{-1}$ in the two-excitation subspace, and therefore become negligible for a large enough number of electrons.

Hence, the Hamiltonian describing the physics of interest of the considered system can be recast in the form
\begin{equation}
    \hat{H}=\hbar\omega_c \hat{a}^\dagger \hat{a}+\sum_{\mu}\epsilon_{\mu}\hat{b}^\dagger_{\mu} \hat{b}_{\mu}
    +\sum_{\mu}\hbar \Omega_{\mu} (\hat{a}^\dagger\ \hat{b}_{\mu}+ \hat{b}_{\mu}^\dagger \hat{a}),
    \label{HamiltonianQuadratic}
\end{equation}
where  $\epsilon_{\mu}=\epsilon_g+\frac{\hbar^2\mu^2}{2m_e^*}$, with $m_e^*$ being the effective electron mass of the conduction band. Note that $\mu$ can be regarded as an effective wavevector along the $z$-direction. However, we do not describe the continuum states as plane waves; instead, as detailed in Appendix~\ref{App:continuum}, we consider the exact continuum single-particle eigenstates of the quantum well. 


\section{Bound polaritons and their quantum  interaction} 
\label{sec:bound}
In order to determine the bound polariton state, namely the excitation created by the interaction of bound-to-continuum intersubband transitions with the cavity photon mode, we have diagonalized exactly the quantum light-matter Hamiltonian in Eq.~(\ref{HamiltonianQuadratic}) in the single-excitation subspace. The polariton state has the form 
\begin{equation}
    \ket{P}=\left[\Phi_{\text{pho}}\hat{a}^{\dagger}+\sum_{\mu}\phi_{\mu} {\hat{b}}_{\mu}^{\dagger}\right]\ket{0}\ket{F},
\end{equation}
where the Hopfield coefficients corresponding to the bound-to-continuum electronic transitions are 
\begin{equation}
    \phi_{\mu}=\Phi_{\text{pho}}\frac{\Omega_{\mu}}{\epsilon_{\mu}-\epsilon_P}.\label{Hopfield}
\end{equation}
The photon Hopfield coefficient is fixed by the normalization condition
$\Phi_{\text{pho}}^2+\Phi_\text{mat}^2=1$, with \begin{equation}\Phi_\text{mat}^2\equiv\sum_{\mu}\phi^2_{\mu}\, ,
\end{equation} being the matter fraction of the polariton excitation. 
\begin{figure}
\begin{centering}
\includegraphics[scale=0.55]{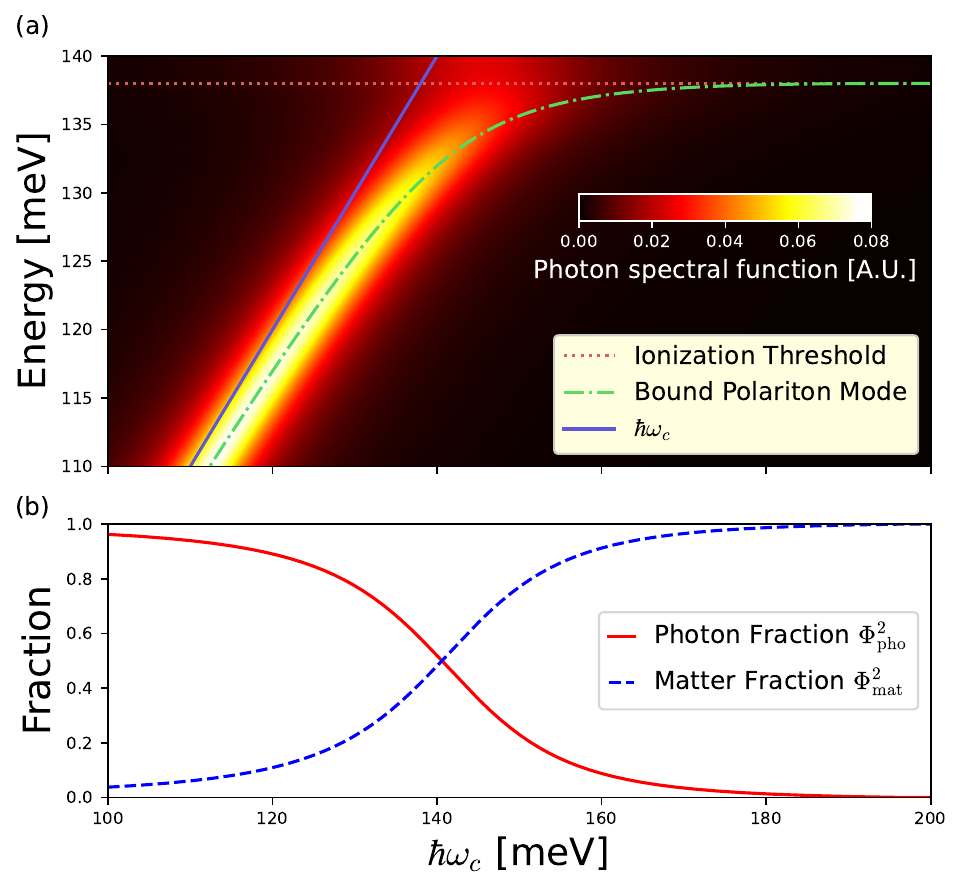}
\par\end{centering}
\caption{ Spectral properties of the bound polariton mode.
(a) Contour plot of the photonic spectral function $A(\epsilon,\hbar\omega_c)$ as a function of the cavity frequency $\omega_c$ and the energy $\epsilon$. To account for cavity losses, a finite imaginary part $\gamma_{\text{c}}$ was added to the self-energy of the Green’s function used to compute the spectral function.
(b) Matter and photonic fractions of the bound polariton as a function of the cavity frequency. The results correspond to a quantum well of width $L_{\text{QW}} = 3$ nm, ionization energy $\epsilon_{\text{ion}} = 138$ meV, effective electron mass $m_e^* = 0.06m_e$, electron density $n = 5\times10^{12}$ cm${^{-2}}$, dielectric constant $\varepsilon = 13\varepsilon_0$, and cavity broadening $\gamma_{\text{c}} = 5$ meV.}
\label{fig:Spectral-function} 
\end{figure}
Illustrative examples of results are shown in Fig.~\ref{fig:Spectral-function},  using the experimental parameters of the experiments reported in Ref.~\cite{Cortese2021}. In particular, panel (a) shows the appearance of the bound polariton state as a resonance in the photonic spectral function, which can be tuned as function of the cavity frequency. The matter and photon fraction of the bound polariton mode are represented in Fig.~\ref{fig:Spectral-function}(b).  These results are in quantitative agreement with the experimental results in  \cite{Cortese2021} and theoretical ones in \cite{Cortese2019} although our approach here is entirely analytical. In another theoretical work  \cite{Kumar2022} the problem was addressed analytically but approximating the continuum states as plane waves. 

Knowing the bound polariton state and tracing out the photonic component, we can calculate the corresponding electron-hole wavefunction. This is given by the expression 
 \begin{equation}
    \Psi(\textbf{r}_\parallel,\textbf{R} _\parallel,z_e,z_h)=\langle \textbf{r}_e,\textbf{r}_h,0|P\rangle  = \psi_e(z_e)\psi_h(z_h)J_0(r_\parallel k_F),\label{eq:WFmatter}
\end{equation}
which is a product of out-of-plane and in-plane wavefunctions.  The Bessel function of first kind $J_0(r_\parallel k_F)$ describes the in-plane electron-hole motion where $r$ is  the modulus of the in-plane electron-hole distance vector $\textbf{r}_\parallel \equiv\textbf{r}_{e\parallel}-\textbf{r}_{h\parallel}$.
\begin{figure}
\includegraphics[scale=0.5]{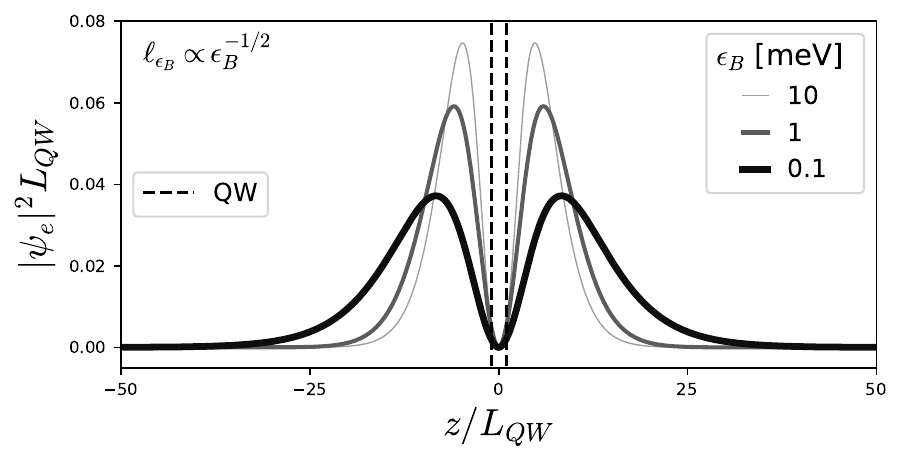}
\caption{Squared wavefunction of the bound polariton state for different values of the polariton binding energy, controlled by the detuning of the cavity mode frequency. Other parameters are the same as in Fig.~\ref{fig:Spectral-function}, which also shows the dependence of $\epsilon_B$ on the cavity detuning.} 
\label{fig:Spread-wave-function}
\end{figure}
The out-of-plane wavefunction is the product of the hole wavefunction, associated with the single quantum-confined subband ($\lambda = 1$) in the quantum well, and an electron wavefunction expressed as a linear combination of continuum single-particle eigenfunctions, namely:
\begin{equation}
    \psi_e(z_e)\equiv \sum_{\mu}\phi_{\mu}\varphi_{\mu}(z_e).
    \label{Psi_e}
\end{equation}
Such wavefunction corresponding to the bound polariton state is depicted in Fig.~\ref{fig:Spread-wave-function} for different values of the polariton binding energy. While the polariton binding energy tends to zero, the spatial spread of the wavefunction dramatically increases.  

It is convenient to introduce a localization length $\ell_{\epsilon_B}$ defined as,
\begin{equation}
  \ell_{\epsilon_B}\equiv \frac{1}{2}\frac{\hbar}{\sqrt{2m_e
  \epsilon_B}},\label{eq:Blength}  
\end{equation}
where the  $\epsilon_B\equiv \epsilon_g-\epsilon_P$ is the polariton binding energy.  

Analogously to the case of standard excitons bound by Coulomb interaction, we can define a Bohr radius as the expectation value of the relative electron-hole distance:
\begin{equation}
    a_B\equiv \int_{-\infty}^{+\infty}\psi^2_h(z_h)\psi^2_e(z_e)|z_e-z_h|dz_e dz_h.\label{eq:BohrR}
\end{equation}
Remarkably, in the limit of vanishing polariton binding energy ($\epsilon_B\to0$), we have that $ a_B /\ell_{\epsilon_B} \to 1 $. 
This analytical result has been confirmed by the numerical evaluation of the integral in Eq.~(\ref{eq:BohrR}), as shown in Fig.~\ref{fig:gPP}(a). As a consequence, the bound polariton's Bohr radius acquires arbitrarily large values close to the ionization threshold, making these excitations reminiscent of Rydberg atoms and excitons. 

\subsection{Polariton-polariton interaction}
\begin{figure}
\begin{centering}
\includegraphics[scale=0.6]{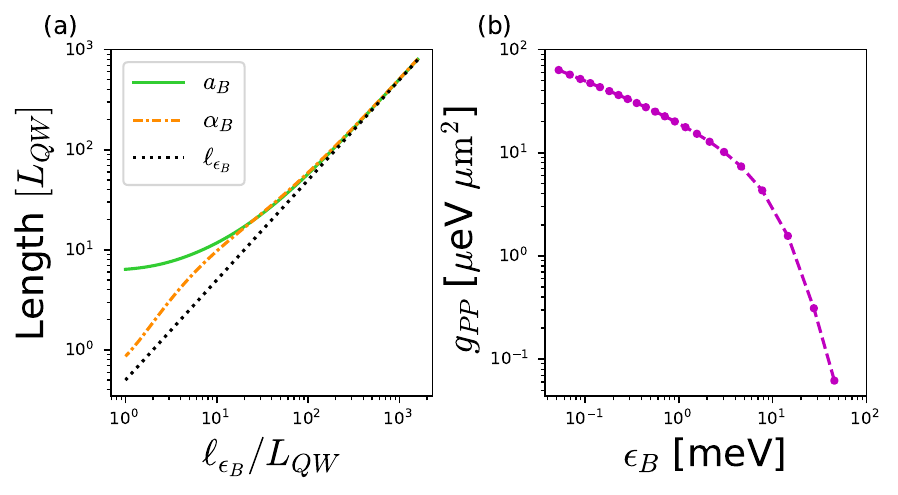}
\par\end{centering}
\caption{\label{fig:gPP}
(a) Dependence of three relevant length scales (electron hole Bohr radius $a_B$, Coulomb scattering length $\alpha_B$, and localization length $\ell_{\epsilon_B}$) on the polariton binding energy $\epsilon_B$. The numerical result (green solid line) is in agreement with our analytical prediction (black dashed line) in the $\epsilon_B \to 0$ limit.
(b) Polariton-polariton interaction $g_{PP}$ as a function of the polariton binding energy, using the experimental parameters from Ref.~\cite{Cortese2021}. The corresponding interaction energy is $U = g_{PP}/S$, where $S$ is the transverse area.
}
\end{figure}

Here we calculate the polariton–polariton interaction exactly at the quantum level. In particular, the quantum anharmonicity is defined as the difference between the energy of two polaritons occupying the same state and twice the energy of a single polariton excitation, namely
\begin{equation}
U \equiv \frac{g_{PP}}{S} \equiv \langle PP | H_B - 2E_P | PP \rangle ,
\end{equation}
where two-polariton normalized~\footnote{For two-polariton states, the adopted normalization includes a correction of order $1/N$, where $N$ is the total number of electrons in the Fermi sea. For two-polariton states, such a correction is negligible when the density of electrons is large enough.} state reads
\begin{equation}
 | PP \rangle  = \frac{1}{\sqrt{2}}\left[\Phi_{\text{pho}} \hat{a}^{\dagger}+\sum_{\mu}\phi_{\mu} \hat{b}_{\mu}^{\dagger}\right]^2 \vert 0 \rangle \vert F \rangle \,.
\end{equation}
Note that the interaction depends on the transverse area $S$ of the two-dimensional system; for this reason, it is convenient to introduce the quantity $g_{PP}$, which has dimensions of energy times area. After some algebra, we get the following result:
\begin{equation}
g_{PP}=\Phi_\text{mat}^4\frac{e^2}{2\epsilon}\alpha_B +\Phi_\text{mat}^3\Phi_{\text{pho}}\frac{\hbar\tilde{\Omega}}{n},\label{Eq:gpp}
\end{equation}
where the scattering length $\alpha_B$, due to Coulomb interaction, is given by the following double spatial integral: 
\begin{align}
\alpha_B=2 \iint |z-z'|\big\{2\psi_e^2(z)\psi^2_h(z')-\psi_h^2(z)\psi^2_h(z')\nonumber\\
+2\psi_e(z')\psi_h(z')\psi_e(z)\psi_h(z)-\psi_e^2(z)\psi^2_e(z')
\big\} dzdz',
\label{AlphaB}
\end{align}
and the effective Rabi frequency reads
\begin{equation}
\tilde{\Omega}=\sum_\mu \frac{\phi_\mu}{\Phi_{\text{mat}}}\Omega_\mu.
\end{equation} 
In Fig.~\ref{fig:gPP}(a), We show both numerically and analytically that the scattering length grows asymptotically as $\alpha_B \sim a_B/2$. This occurs in the regime of vanishing polariton binding energy ($\epsilon_B\to0$), that is when the Bohr radius becomes much larger than the quantum well width ($a_B\gg L_{QW}$).  In Fig.~\ref{fig:gPP}(b) we report the polariton-polariton interaction using the same experimental parameters used in~\cite{Cortese2021}. Our result predicts dramatically enhanced polariton-polariton interactions when the binding energy approaches zero. In this limit, the Coulomb scattering term in Eq.~(\ref{Eq:gpp}) dominates over the saturation nonlinearity \cite{Zanotto2012,Nguyen-The2013} leading to a strong polariton-polariton repulsion.  

We would like to point out that if one is interested in the resonant optical excitation of the considered polaritons, a compromise must be found when tuning the cavity mode frequency in order to achieve both large quantum nonlinearities and energy selectivity. Indeed, an excessively small binding energy would make resonant optical excitation unfeasible. Depending on the applications and the type of system, there will be a sweet-spot detuning where a small but finite binding energy allows for both resonant excitation and enhanced nonlinearities.

\subsection{Comparison with bound-to-bound transitions}
Our theoretical framework can also be applied to standard intersubband polaritons arising from strong cavity coupling to bound-to-bound intersubband transitions. In this case, one must simply use the corresponding Hopfield coefficients (photonic and matter fractions) and  the single-particle electron wavefunction corresponding to the second quantum-confined subband in a quantum well, rather than the cavity-induced electronic wave function in Eq.~(\ref{Psi_e}), to compute the scattering length $\alpha_B$ and the Rabi frequency [Eq.~(\ref{RabiFreq})].

As an illustrative example, let us consider the intersubband transition between the two lowest conduction subbands in a quantum well with infinite barrier height. The corresponding scattering length can be obtained analytically as
\begin{equation}
\alpha_B^{\text{1-2}}=-\frac{275}{72\pi^2}L_{QW} \simeq - 0.39 L_{QW}.
\end{equation}
Hence, for bound-to-bound intersubband polaritons, if we consider the lower branch, the Coulomb interaction term proportional to $\alpha_B$ in Eq.~(\ref{Eq:gpp}) becomes negative (attractive), in agreement with previous studies~\cite{Nguyen-The2013,Knorr2022}. In this regime, there is a partial compensation between the attractive Coulomb term (proportional to $\alpha_B$) and the repulsive second term in Eq. (\ref{Eq:gpp}) arising from the saturation nonlinearity of the vacuum Rabi frequency due to Pauli blocking. Equally importantly, the associated scattering length is smaller than the quantum well width. For the same system parameters (quantum well width, dielectric constant, and effective electron mass), we find $g_{PP}^{\text{1-2}}\approx -0.20~\mu\text{eV}~\mu\text{m}^2$ at zero frequency detuning with the cavity mode. This interaction strength is significantly smaller, by approximately two orders of magnitude, than the values reported in Fig.~\ref{fig:gPP} for the case of bound-to-continuum intersubband transitions.

\section{Conclusions and perspectives}
\label{sec:conclusions}
In this work, we have presented the first investigation of interactions between polaritons arising from the coupling of a cavity photon mode to bound-to-continuum intersubband transitions in doped quantum wells. We have shown that these excitations—excitons bound through photon exchange, recently demonstrated experimentally \cite{Cortese2021}—exhibit strongly enhanced polariton–polariton interactions reminiscent of those in Rydberg atoms. In particular, the polariton binding energy can be tuned by the cavity-mode frequency, and the interaction strength increases dramatically as the binding energy is reduced. In this regime, the spatial extent (Bohr radius and associated electric dipole) of these excitations becomes much larger than the quantum well width. The resulting quantum nonlinearities are therefore much stronger than in standard bound-to-bound intersubband polaritons \cite{Dini2003, Nguyen-The2013, Knorr2022}, where the electric dipole is limited by the quantum well width. Moreover, an additional appealing feature of the bound polaritons formed through bound-to-continuum intersubband transitions is that the dark intersubband electron–hole excitations are are in the continuum and delocalized. The enhanced quantum optical nonlinearities of these polaritons could be highly relevant for both fundamental studies and quantum optoelectronic applications in the mid- and far-infrared.

\acknowledgements
This work has received support under the program "Investissement d'Avenir," launched by the French Government and implemented by the ANR, with the reference "ANR‐22‐CMAS-0001, QuanTEdu-France," and from the ANR grant UNIPOLASER (ANR-23-CE24-0007). We also acknowledge support from the grant Polaritonic, funded by the French Government and managed by the ANR under the France 2030 programme, with the reference ANR-24-RRII-0001. We would like to thank R. Colombelli, M. Jeannin and J. M. Manceau for fruitful discussions.

\appendix
\section{Continuum states}
\label{App:continuum}
In this Appendix, we provide details on the continuum single-particle eigenstates. As discussed in the main text, we consider a quantum well supporting only one quantum-confined state. For simplicity, we express the wavefunction as a function of the dimensionless out-of-plane coordinate $\zeta = 2z/L_{QW}$, where $L_{QW}$ is the spatial width of the quantum well.

The single-particle quantum-confined state, where the hole is created, has the wavefunction
\begin{align}
    \varphi_{1}(\zeta)= \phi_{h}(\zeta) =\frac{1}{\sqrt{L_{QW}}} \sqrt{\frac{\kappa}{\kappa+\cot{\kappa}}}\times \nonumber\\  \begin{cases}
        \cos \zeta \kappa & |\zeta|\leq 1,\\
        e^{-\kappa |\zeta -1|\tan\kappa}\cos\kappa & |\zeta|> 1, 
    \end{cases}
\end{align}
where $\kappa$ is the quantum number determined by the quantization condition $\kappa \tan\kappa=\sqrt{\kappa ^2+\delta^2}$, with $ \delta=L_{QW}\sqrt{2mV_0\hbar^{-2}}$ and $V_0$ the potential barrier height of the well.

For the unbound states with energy $E$, we consider a system of size $L$ along the $z$-direction. We label these states by $\mu =L_{QW}\sqrt{2mE\hbar^{-2}}$. We also define the auxiliary quantity $\mu'=\sqrt{\mu^2+\delta^2}$. The unbound single-particle wavefunctions are given by the following expression:
\begin{align}
    \varphi_{\mu}(\zeta)= \frac{1}{\sqrt{\pi L}}\frac{1}{\sqrt{1+\frac{\delta^2}{\mu^2}\cos^2\mu'}}\times\nonumber\\
    \begin{cases}
        \sin{\mu'z}& {\rm for} \,|z| \le  1\\
        \frac{\mu'}{\mu}\sin(|z-1|\mu)\cos\mu'+\cos(|z-1|\mu)\sin\mu' &  {\rm for} \,|z| >  1  . \end{cases}
\end{align}

In the dipolar gauge~\cite{Todorov2012}, the vacuum Rabi coupling associated with the bound-to-continuum transition (Eq.~\ref{RabiFreq}) is proportional to the dipolar matrix element defined as
\begin{equation}
    z_{1\mu}=\int z \,\phi_h(z)\varphi_{\mu}(z)dz.
    \label{z1mu}
\end{equation}
Note that in the limit $L \to + \infty$, the energy spectrum becomes continuous, and all quantities related to the bound-to-continuum polariton become independent of $L$.

\section{Photonic spectral function}
\label{App:photonic}
The photonic spectral function $A(\epsilon)$ is defined in terms of the photon's retarded Green's function $G_p(\epsilon) $ via the following expression:
\begin{equation}
A(\epsilon)=-2\text{Im}G_p(\epsilon).
\end{equation}
The photon and photon-matter retarded Green's functions are defined in the time domain respectively as 

\begin{align}
    \tilde{G}_p(t)&=-i\theta(t)\langle[\hat{a},\hat{a}^\dagger(t)]\rangle_{GS},\\
    \tilde{G}_\mu(t)&=-i\theta(t)\langle[\hat{a},\hat{b}_\mu^\dagger(t)]\rangle_{GS},
\end{align}
where $\theta(t)$ is the Heaviside function and $\langle[\dots]\rangle_{GS}$ represents the expectation value over the ground state. In order to obtain the photonic $G_p(\epsilon) $, we have determined the equations of motion for the coupled Green's function following the method in Ref.~\cite{Bruus2004} .  The coupled equations read:
\begin{align}
    (\epsilon-\hbar\omega_c)G_p(\epsilon)-\sum_\mu \hbar\Omega_\mu G_\mu(\epsilon)&=1,\\
    (\epsilon-\epsilon_\mu)G_\mu(\epsilon)-\hbar\Omega_\mu G_p(\epsilon)&=0.
\end{align}
The solution for the photonic Green's function is:
\begin{equation}
    G_p(\epsilon+i\eta)=\frac{1}{\epsilon- \hbar\omega_c-\Sigma(\epsilon)+i\eta}.
\end{equation}
The photon's self-energy is defined as 
\begin{equation}
    \Sigma(\epsilon)=-\sum_\mu \frac{\hbar^2\Omega_\mu^2}{\epsilon_g-\epsilon+\frac{\h^2\mu^2}{2m_e}}.
\end{equation}
Taking the continuum limit,  we can evaluate the related self-energy with the following integral
\begin{equation}
    \Sigma(\epsilon)=-\int_{-\infty}^{+\infty}\frac{\hbar^2\bar{\Omega}^2_\mu }{\epsilon_g-\epsilon+\frac{\hbar^2\mu^2}{2m_e}}\frac{d\mu}{2\pi},
\end{equation}
where  $\bar{\Omega}_\lambda=\Omega_\mu\sqrt{L}$.  Note that this quantity is well defined in the continuum limit since $\Omega_\mu\propto \sqrt{L}^{-1}$ as it can be seen by inspecting the dipolar matrix element definition in Eq.~(\ref{z1mu}).

The contour plot of the photon spectral function depicted in Fig.~\ref{fig:Spectral-function}  has been obtained with the explicit expressions reported in this Appendix. In particular, the bound polariton energy is found from the pole of the full photonic retarded Green's function and used to compute the Hopfield coefficients given by Eq.~(\ref{Hopfield}).

\bibliography{polaritons_ref}

\newpage
\onecolumngrid
\end{document}